\documentclass[preprint]{aastex}

\newcommand{\bs}{$\backslash$}
\newcommand{\ww}{WW C{\sc ygni}}
\newcommand{\bul}{$\bullet$}

\shorttitle{Period changes of WW Cygni}
\shortauthors{Zavala et al.}

\begin{document}

\title{The origin of cyclic period changes in close binaries: the case of \\
    the Algol binary WW Cygni}

\author{Robert T. Zavala\altaffilmark{1,2}, Bernard J. 
McNamara\altaffilmark{3}
, Thomas E. Harrison\altaffilmark{4}, Eduardo Galvan, Javier Galvan, Thomas Jarvis, 
GeeAnn Killgore, Omar R. Mireles, Diana Olivares, Brian A. Rodriguez, Matthew Sanchez,  
Allison L. Silva, Andrea L. Silva, Elena Silva-Velarde}
\affil{Department of Astronomy, New Mexico State University,
    Dept 4500 P.~O. Box 30001, Las Cruces, NM  88011}

\altaffiltext{1}{rzavala@nrao.edu}
\altaffiltext{2}{Present address: National Radio Astronomy Observatory,
P.O. Box 0, Socorro, NM 87801}
\altaffiltext{3}{bmcnamar@nmsu.edu}
\altaffiltext{4}{tharriso@nmsu.edu}

\begin{abstract}
Years to decade-long cyclic orbital period changes have been observed in several classes
of close binary systems including Algols, W Ursae Majoris and RS Canum Venaticorum
systems, and the cataclysmic variables. The origin of these changes
is unknown, but mass loss, apsidal motion, magnetic activity, and the presence
of a third body have all been proposed. In this paper we use new CCD
observations and the century-long historical record of the times of primary eclipse
for WW Cygni to explore the cause of these period changes. WW Cygni is an Algol binary
whose orbital period undergoes a 56 year cyclic variation with an amplitude of
$\approx 0.02$ days.
We consider and reject the hypotheses of mass transfer, mass loss, apsidal motion
and the gravitational influence of an unseen companion as the cause for these changes.
A model proposed by Applegate, which invokes changes in the gravitational quadrupole
moment of the convective and rotating secondary star,
is the most likely explanation of this star's orbital period changes.
This finding is based on an examination of WW Cygni's residual O$-$C curve and an analysis
of the period changes seen in 66 other Algols.  Variations in the gravitational quadrupole
moment are also considered to be the most likely explanation for the cyclic
period changes observed in several different types of binary systems.
\end{abstract}


\section{Introduction}

Long term cyclic period changes are a fairly common phenomenon in close binary systems.
These changes have been observed in Algols, RS Canum Venaticorum and W Ursae Majoris binaries,
and they are suspected to be present in Cataclysmic Variables (CVs)( Hall 1989; Hall \& 
Kreiner 1980; Hobart et al. 1994; Warner 1988). Researchers have found it difficult to 
quantify this phenomenon because precise measurements
extending over decades are required. The traditional manner in which period changes 
are investigated is through 
the construction of a plot of the difference between a binary's ${Observed}$ and ${Calculated}$
times of primary eclipse versus time. This plot is referred to as an O$-$C diagram.
\citet{lan99} estimate that the median modulation period seen in these diagrams is 
40-50 years for Algol and RS CVn systems while for CVs this value varies from years to decades.
The amplitudes of these waveforms are also small having $\Delta\rm{P/P} = 10^{-7} - 10^{-6}$ for 
CVs, $\Delta\rm{P/P} = 10^{-6}$ for W UMa binaries, and $\Delta\rm{P/P} = 10^{-5}$  Algols and 
RS CVns.  Although the physical cause for these variations is not known, the similar appearance of 
the O$-$C diagrams for several classes of close binaries suggests that a common 
mechanism produces them.

Several hypotheses have been advanced to explain the cyclic period changes in close 
binaries. Apsidal motion, which involves a change in the orientation of the binary's major
axis, is an unlikely mechanism because close binaries possess circular orbits and
this phenomenon only occurs in systems having large eccentricities. If apsidal motion were 
present, the times for secondary and primary minima would be shifted in opposite 
directions, but this effect is rarely seen and does not appear to be present in Algols.  
An alternate possibility 
is that the cyclic pattern in the O$-$C plot is caused by the presence of a third
body. This idea has been explored by several investigators, most recently by \citet{bheg96}. 
These researchers postulate that the motion of the binary around the center of mass of a triple
system causes the primary and secondary eclipse times to vary in a uniform and periodic fashion. 
In this case, the cyclic O$-$C pattern arises from orbital induced changes in the distance to the
observer.  Finally, \citet{hall89} noted that, for Algols which
possess alternating period increases and decreases, the spectral types of the secondaries are
always in the range late F to K. These stars have outer convective zones and, if they are rapidly
rotating, a magnetic dynamo is produced \citep{par79}. \citet{App92} and \citet{Lan98} used this 
information to 
develop a theory to explain the cyclic shape of the O$-$C curves of these systems. They
suggest that if the secondary is deformed by tidal and centrifugal forces, changes in the
internal rotation associated with a magnetic activity cycle alter the star's gravitational
quadrupole moment.  As the quadrupole moment increases the gravitational field increases,
leading to a decrease in the binary period.  Conversely, when the quadrupole moment
decreases, the binary period increases \citep{lan99}.  

In this paper we will use an extraordinarily long O$-$C curve for WW Cygni to examine whether the 
above models can explain the cyclic pattern of period variations observed in this object.
In \S 2 we present our new observations of WW Cyg. In \S 3 an updated ephemeris is computed and
residuals in the O$-$C plot are discussed and in \S 4 the three models above are tested.  
Conclusions are presented in \S 5.

\section{Program Object and Observations}

WW Cygni (HD 227457, BD+41 3595, HIP 98814)
is a relatively bright
Algol binary with deep ($\Delta$v = 3.5mag) primary eclipses. WW Cyg's visual magnitude
varies from 10.0 to 13.5 with a period of 3.3 days. This system's brightness and
large amplitude have made it a frequent source of study with
over a century of observations present in the literature (see Hall \& Wawrukiewicz 
1972 and references therein). \citet{Str46} established the spectral type of the primary
(mass-gaining) star as B8 and speculated that the secondary (mass-losing) star was
type G. \citet{yoon94} subsequently assigned the secondary a spectral class of G9,
but due to a paucity of sub-giant calibration stars they were unable to determine its
luminosity class. They also noted that photometrically determined spectral types
for Algol secondaries are typically later than those determined spectroscopically.

Multicolor CCD observations of WW Cyg were acquired at the Clyde Tombaugh Campus Observatory 
on the campus of New Mexico State University between October 1997 and October 1999.
The observatory is equipped with a 16 inch F/10 Meade LX 200 Schmidt-Cassegrain
telescope and a Santa Barbara Instruments Group ST8 CCD camera. The camera
uses a KAF1600 CCD with 1520$\times$1020 9$\mu$m pixels. Readnoise and gain were
experimentally determined to be 11 $\rm e^{-}pix^{-1}$ and 2.5 $\rm e^{-}ADU^{-1}$ 
\citep{maj92}. The camera is equipped with UBVRI filters which match the
prescription of \citet{bess90}.
Due to the system's poor sensitivity in U, no observations were collected in
this filter. Because the chip plate scale is 0.46 arc seconds/pixel and typical seeing
at the observatory is 2.5-3'', pixels were binned by two prior to readout. This
procedure decreased the normal readout time from one minute to 15 seconds. Images were
stored on a laptop computer whose internal clock was set at the beginning of each
night using a UTC time signal from a GOES weather satellite. Occasional checks of the
accuracy of the clock were made during an observing run  by initiating exposures
synchronously with the WWVB time signal received via shortwave.
Exposures were 120 seconds in B, 60 seconds in V, and 30 seconds in R and I. On two
occasions (UT 1998 September 01 and October 11) observations were acquired in V only to reduce
the sampling time interval so that a more precise determination of the time of primary 
minimum could be obtained. Otherwise, with a few exceptions, the four filters were cycled 
in the order BVRI.  Data were reduced using the aperture photometry packages found in 
IRAF\footnote{IRAF is distributed by the National Optical Astronomy Observatories,
which are operated by the Association of Universities for Research
in Astronomy, Inc., under cooperative agreement with the National
Science Foundation.}
and standard differential photometry techniques. The B solution 
was poor and therefore was not used. Magnitudes for WW Cyg were obtained 
relative to two comparison stars from the HST Guide Star Catalog: GSC 3158-1468 (V=12.08 and 
V$-$I=0.66) and 3158-1228 (V=11.51 and V$-$I=1.74). The complete multi-color data
will be presented in a later paper. 

\section{Ephemeris and the O$-$C Curve}
By definition, Algol systems involve mass transfer and therefore period changes are
expected to occur \citep{kvw58}. To investigate these changes a new ephemeris was
calculated for WW Cyg and its century-long O$-$C curve was reexamined.
Observations near primary eclipse were obtained at NMSU on UT 1998 August 22, September 01, and
October 11. The light curves for these eclipses are shown in Figure 1, along with the
comparison star data. Preliminary times of minima (TOM) were then determined from a
visual inspection of these observations. These times were then refined
using the algorithm of \citet{kvw56} and a computer code kindly provided by
Dr. R. Nelson (2000, private communication) (see also Mallama 1982). 
The new TOM are listed in Table 1
along with an earlier TOM determined
by \citet{buck98}.  A weighted linear least squares fit to these TOM gives the
following ephemeris
\begin{equation}
HJD = 2,450,387.6071 \pm 0.0005 + (3.317813 \pm 0.000003)E.
\end{equation}

\citet{haw72} conducted a thorough literature search and produced an O$-$C plot of
WW Cyg through 1972 using the ephemeris of \citet{gra22}

\begin{equation}
HJD = 2,416,981.3134 + 3.317676E.
\end{equation}

\noindent In Table 2 we list post 1972 O$-$C data with respect to Graff's ephemeris.
This allows us to extend the O$-$C plot by almost 30 years. We plot the entire 107 year
O$-$C data in Figure 2 together with the best fit parabola,

\begin{equation}
C = 0.012 - 7.78\times10^{-6}E + 6.92\times10^{-9}E^2,
\end{equation}

\noindent with all points weighted equally.

\section{Discussion of the WW Cyg O$-$C Curve}
The long term trend in Figure 2 indicates an increasing period. Assuming this trend is
the result of mass transfer we attempted an estimate of the mass transfer rate
$\dot{M}$. \citet{kvw58} derived a
relationship for the change in the orbital period assuming total orbital 
angular momentum is conserved while mass is transferred from one star
to another. Reproducing equation 5 of \citet{kvw58}:

\begin{equation}
{\Delta P}/P = 3(m_{s}/m_{p} - 1)dm_{s}/m_{s},
\end{equation}
 
\noindent where ${\Delta P}$ is the change in the period ${\it P}$, $m_{s}$ is the mass of the 
secondary, $m_{p}$ is the mass of the 
primary, and $dm_{s}$ is the change in mass of $m_{s}$.
An estimate for the mass transfer rate in WW Cyg can now be made. Between the 
ephemeris of \citet{gra22} and equation 1 the period increased by 0.0002 days. 
Using the solution for WW Cyg presented by \citet{haw72} the mass of the secondary can be 
set to 2$\rm M_\odot$ and the mass ratio to $\approx 0.4$. Over an elapsed time of 
100 years equation 4 gives as a mass transfer estimate

\begin{equation}
\dot{M} = 7 \times 10^{-7} M_{\odot}yr^{-1}.
\end{equation}

\noindent This simplistic treatment of mass transfer is probably an underestimate of the actual 
mass transfer rate. A more realistic model for non-conservative mass transfer is the wind-driven 
mass-transfer model derived by \citet{th91}. The mass transfer rate determined for 
the Algol U Cep in \citet{th91} is an order of magnitude larger than the estimate above.

A careful examination of the O$-$C curve in Figure 2 shows that WW Cyg possesses an 
alternating sequence
of period increases and decreases superimposed upon a parabola. This behavior
is more easily seen in Figure 3 where the residuals from the parabolic fit are shown.
Below we examine three of the most widely discussed hypotheses used to explain this type of
cyclic behavior.

{\it a) Mass exchange or mass loss due to stellar winds}: The most frequently mentioned explanation
for period changes in close binary systems is the transfer of mass from one star to another.
However, mass transfer from a less massive to a more massive component 
results in a steadily increasing period not an alternating
sequence of period increases and decreases.  Mass transfer also appears to be an unsuitable 
explanation for the cyclic O$-$C pattern seen in close binaries for another reason.  The RS CVn 
systems, in which cyclic O$-$C patterns are present, are in general detached systems 
\citep{hall76}, so mass transfer cannot be invoked.   
Mass loss due to stellar winds also appears to be ruled out. \citet{deb79} have shown that 
for RS CVn stars the mass loss rates required to produce a quasi-periodic signal in the O$-$C
diagram are orders of magnitude larger than allowed by observations. They note that such 
losses would conflict with the detection of soft x-rays from $\rm \alpha$ Aur, UX Ari, 
HR 1099, and RS CVn.  If we assume a single process produces the cyclic O$-$C curves, 
neither of these processes appear to be viable.

{\it b) The Presence of a Third Body}: The possibility that the cyclic O$-$C changes
in Algols and other short period binaries are due to the presence of a third star has been 
discussed by \citet{fch73}, \citet{cham92}, and recently by \citet{bheg96}.  In this model the 
binary revolves around the center of mass of the system thereby creating a regular
change in the observed period due to a light travel time effect (LTTE). Borkovits and
Heged$\rm\ddot{u}$s tested this suggestion by incorporating the gravitational effects of
a third body into fits of the residual O$-$C curves of 18 close binaries. This model was able to 
produce qualitative agreement between the observed and computed O$-$C curves for 4 systems 
(1 Algol, 3 W UMa) by solving for the orbital elements of the assumed third body.
In general, the mass of the third star was less than 0.5 $\rm M_{\odot}$ (with one exception)
and therefore could have escaped detection.
Marginal evidence for a third body was presented for four other
systems (all Algols). Marginal solutions required more than one unseen companion and the mass for 
these companions, more than 2 $\rm M_{\odot}$ depending on eccentricity, begins to present 
non-detection problems. For eight other systems LTTE solutions of dubious reliability were obtained
by Borkovits and Heged$\rm\ddot{u}$s for the sake of completeness. LTTE solutions could not be found 
for only two of the 18 binaries they studied. Motivated by this study we used the O$-$C data for WW Cyg 
shown in Figure 2 to further 
test this hypothesis.

\citet{kop78} has shown that a third object orbiting a close binary creates a periodic pattern
in an O$-$C curve and that by describing this signal as a Fourier series, the orbital
parameters of the third body can be determined.  Following the method adopted by \citet{bheg96},
we first subtracted the best-fit parabola shown in Figure 2 from the WW Cyg data.
The power spectra of these residuals were then computed and yielded a waveform having
the parameters

\begin{equation}
R = 0.0187cos[2\pi(1.61\times10^{-4})E - 1.2712],
\end{equation}

\noindent where $R$ is the waveform of the residuals in days, $E$ is the epoch,
1.61 $\times10^{-4}$ is the inverse period in epochs of Graff's ephemeris, and the phase is 
$-1.27$ radians.
This power spectrum is shown in Figure 4 as a solid line. When
equation 4 is plotted on the actual data (Figure 5a), it did not provide a satisfactory fit.
Consider the residuals in two groups; one from 1890 to 1940 and the other from 1940
to 1998. The earlier group appears to have a narrow inverted V shape, and a shorter period
compared to the later group. Additionally, the observations of 1890-1900 and 1960-1972
deviate from the predicted form.

If the third body possesses an orbit with a nonzero eccentricity, the waveform
contains both fundamental and first harmonic terms.  Adopting equation 6 as the fundamental, 
we searched for its first harmonic at twice the frequency of the fundamental.  The fundamental 
was subtracted from the data and the 
power spectrum was recomputed, but the subsequent power spectrum was found to be noise 
dominated (see lower dashed curve
in Figure 4). We then used an iterative non-linear least squares fitting routine, based on the 
Marquardt 
algorithm \citep{pre92} to determine the parameters of the first harmonic. This program seeks 
to identify individual coefficients by varying designated input values so that the sum of the 
squares 
of the residuals are minimized. As an initial guess, an amplitude of 10 percent of the
fundamental was employed since this value was similar to that found by \citet{bheg96} for
other close binaries. The parameters of the fundamental (amplitude, frequency and phase) 
were fixed as was the frequency of the first harmonic. 
According to \citet{kop78}, the eccentricity of the third companion
is given as

\begin{equation}
e = 2\sqrt{\frac{a_2^2 + b_2^2}{a_1^2 + b_1^2}}
\end{equation}

\noindent where $e$ is the eccentricity of the orbit and the $a_i$'s and $b_i$'s are the
coefficients of the cosine and sine terms respectively of the Fourier components of the O$-$C
residuals. Our assumed amplitude yields an eccentricity of $\sim$ 0.1.
With the amplitude and phase of the first harmonic as the only free parameters, the fitting 
routine consistently produced unrealistic eccentricities
that were greater than one. Fixing only the frequencies and leaving the amplitudes and phases
of the two waveforms as free parameters produced unrealistic fits to the data. Constraining the 
amplitude of the first harmonic 
at 0.002 days and leaving the phase 
of the first harmonic as the only free parameter resulted in an unacceptably
high reduced $\chi^{2}$ of greater than 130.  Based on these experiments, we conclude that
the third body hypothesis fails to explain the observed O$-$C curve for WW Cyg.
This agrees with the naive expectation from Figure 5a as the residuals do not appear to be
strictly periodic.

{\it c) Magnetic cycles}: In an important study comparing RS CVn to Algol binaries, \citet{hall89}
found a striking correlation between the development of surface convection in the low mass 
secondary and the presence of an alternating orbital period. Hall's plot, reproduced here as
Figure 6, illustrates that when the secondary has a spectral type earlier than about F5 the 
orbital period never displays an alternating pattern of period increases and decreases. 
For systems where the secondary has a spectral type later that an F5, a significant number of 
binaries possess alternating periods.
Hall interpreted this finding as evidence that the onset of convection
plays an important role in producing cyclic period changes. \citet{war88} and \citet{App87} 
attempted to explain the O$-$C patterns by assuming deformations of 
the star away from hydrostatic equilibrium due to tidal or magnetic pressure effects, 
respectively. However, \citet{Mar90} showed that 
tidal or magnetic pressure induced distortions in the active star away from hydrostatic 
equilibrium were ruled out based on energetic considerations. Such distortions could not
be the method by which the quadrupole moment is deformed.  

\citet{App92} and \citet{Lan98} subsequently proposed that a cyclic O$-$C pattern could be 
produced if the active star's internal angular momentum distribution changes as the star 
goes through a magnetic activity cycle similar to that of the Sun. \citet{App92} avoids the
 energy quandary mentioned above by maintaining fluid hydrostatic equilibrium throughout the 
cycle. This was suggested as a common mechanism to explain the same alternating period changes 
seen  in W UMas, RS CVns, and CVs. This suggestion is consistent with the observational 
evidence for RS CVn-like magnetic activity signatures in Algols (Richards 1990; 
Richards \& Albright 1993). 

In the models of \citet{App92} and \citet{Lan98}, the rotational oblateness of the late type 
star produces a change in the gravitational quadrupole moment, and hence the orbit.
In both models, the gravitational field of the primary is treated 
as originating from a point mass. The field of the secondary is computed as arising from
a point source located at the star's barycenter and a quadrupole moment term caused by
tidal and rotational deformations.  Changes in the star's effective angular
velocity are then directly related to changes in the quadrupole moment and to changes in the
orbital period ( $\rm |\Delta \Omega /\Omega| \sim |\Delta Q| \sim |\Delta P/P|$).
The angular velocity changes as a result of a torque supplied by the magnetic field 
in the outer convective region of the late-type star.  
The presence of the required dynamo in the convective secondary stars is consistent with
the results of \citet{lan99}. Their analysis showed that the modulation period of 46 
close binaries supported the view that a dynamo operated in the outer atmospheres of the 
secondary stars. The principal advantages of this model are that it presents a physically 
plausible explanation whose time scale is in rough agreement with that seen in the O$-$C 
plots and that it removes the need for strict periodicity.

Figure 6 demonstrates that there is no correlation of orbital period changes with mass ratio. 
However, if the gravitational quadrupole 
moment term plays a dominant role in the creation of cyclic
period changes, a correlation may also exist between the occurrence of an
alternating period change and a binary's orbital separation. Cyclic O$-$C variations
should be more pronounced in systems where the secondary star is most deformed, i.e.
when the orbital separations are small.  Conversely, when the separations are large, the
secondary star should not be as deformed and cyclic variations should not be as 
pronounced.
To test this idea we constructed Figure 7 which shows whether these predictions
are supported by the available data. As Hall did in Figure 6 we selected Algols with
convective secondaries from \citet{giu83}. We then searched the literature for
published O$-$C curves for these systems and present the data in Table 3. 
The columns in this table refer to the systems we selected,
the secondary spectral type, the binary period in days, nature of the period change, and the 
reference(s) for the O$-$C data.
In a few cases the secondary spectral type in Table 3 differs from that in  \citet{giu83}. 
The reference for the updated secondary spectral type is given as a footnote for these systems. 
The form of 
the period change noted in Table 3 follows the convention used in Figure 6 with the same 
symbols being employed in Figure 7 for consistency. It is clear from Figure 7 that 
alternating period 
variations 
are prominent in binaries with a period under 6 days and that alternating period variations
are less common in systems with larger orbital separations. 
We consider this result to be suggestive, however more data is needed to confirm it. This 
is particularly true for those systems not marked by an ``X'' 
which possess short periods. 
We note that the low number of Algols with periods greater than 10 days is also insufficient 
to make a firm statement about the nature of the O$-$C plots in this region of the diagram. 
Observational selection effects probably
play a role here because eclipse timings for these systems are more difficult to obtain. 
Additionally, \citet{giu83} show that almost 80 percent of their 101 Algols have periods
under 6.5 days.

Additional predictions of the Applegate model are that the brightness of the active star should
vary with the same period as seen in the O$-$C plot and that this change should be about 0.1 
magnitude.  Based on arguments about the manner in which kinetic and magnetic energy are
exchanged, \citet{lan99} suggested that 0.1 magnitude is probably an 
upper limit. Because the brightness change is caused by variations in the star's differential 
rotation and not to changes in its radius, the star should also be bluer when at maximum 
brightness. 

These predictions were tested by Hall (1991) using the RS CVn system CG Cyg. The active star 
in this system has a spectral type of G9.5V. Hall found that this system's O$-$C curve 
possessed a cyclic pattern with a period of about 52 years (see Figure 1 in Hall 1991). 
Using plots of this object's
mean brightness and color while outside eclipse, he found that these quantities
varied with the same period as the O$-$C curve. He also found that this system was bluest when 
it was brightest. Although Hall argued that these results confirmed 
the Applegate model, the magnitude and color data he presented covered only about
half of the 52 year cycle. 
This conclusion must therefore be considered as tentative.  Nevertheless, if Hall's result 
is accepted, it suggests that, like the Sun, the surface 
layer of the active star is spinning faster than its lower layers. We can perform a similar, but 
cruder, analysis for WW Cyg.  When \citet{haw72} observed this star on JD = 2440407.7 
(Epoch 7061) it had a primary minimum of V=13.26 and dereddened colors appropriate to
a G2III or K0V. When we observed WW Cyg 
on JD= 2451057.8 (Epoch 10,268) it had V=13.46 and dereddened V$-$R appropriate to
a K2V. 
Examining Figure 5a, we found that the times when the secondary was at its brightest(bluest) 
and faintest(reddest) were when the O$-$C wave was at its maximum and minimum values, 
respectively. According to Applegate's (1992) model this implies that
the surface layer of the secondary of WW Cyg is rotating more slowly than its subsurface layer. 
This result differs with that found by Hall for the G9.5V star 
in CG Cyg, which has a similar spectral type. This difference in rotational structure 
might appear to be troubling but the secondary in WW Cyg is most likely 
a subgiant so this example does not provide a direct comparison. Nevertheless, it does illustrate 
how long term period changes due to magnetic activity may provide a probe of the interior
structures of late-type stars in close binary systems.

\section{Conclusions}

WW Cyg, like several other close binaries, displays an alternating pattern in
its times for primary eclipse.  We have explored three possible causes for this cyclic
behavior: mass exchange and mass loss due to stellar winds, the presence of a third body, and
magnetic cycles. The first two possibilities appear to be ruled out. However, the current
evidence is consistent with the idea that magnetic cycles within a lower mass secondary
of spectral type later than F5, play a role. Convection and rotation leading to
magnetic dynamo activity have been theoretically shown to produce cyclic changes in the
gravitational field of the system. These changes are of sufficient strength to produce orbital 
period changes of the observed size in close binaries. Additionally, the time frame for these
period changes is of the same order of magnitude as the magnetic cycles of late 
type stars \citep{bal95}. Although we feel
that the evidence at this point is not conclusive, variations in the gravitational quadrupole
moment caused by magnetic activity appears to provide the best explanation for the
cyclic behavior of the O$-$C curves observed in WW Cyg and other close binary systems.
More data are needed to construct a larger sample of O$-$C curves for close binaries. 
We urge that both magnitude and color information be collected for a variety of 
close binaries to show how the internal rotation of the late-type star changes with time.  
Because several of these systems are relatively bright, this type of observational program is
well suited to facilities equipped with telescopes of modest aperture.

\acknowledgments
We are indebted to the many observers, amateur and professional, who amassed the wealth of data 
on the eclipsing binaries listed in Table 3. This research has made use of the SIMBAD database, 
operated at CDS, Strasbourg, France. This research has made use of NASA's Astrophysics
Data System Abstract Service. NSF Grant HRD-9628730 supported this research. RTZ gratefully
acknowledges support from the New Mexico Alliance for Graduate Education and the Professiorate 
through NSF Grant HRD-0086701.

\clearpage

\begin{figure}
\vspace{19.2cm}
\includegraphics{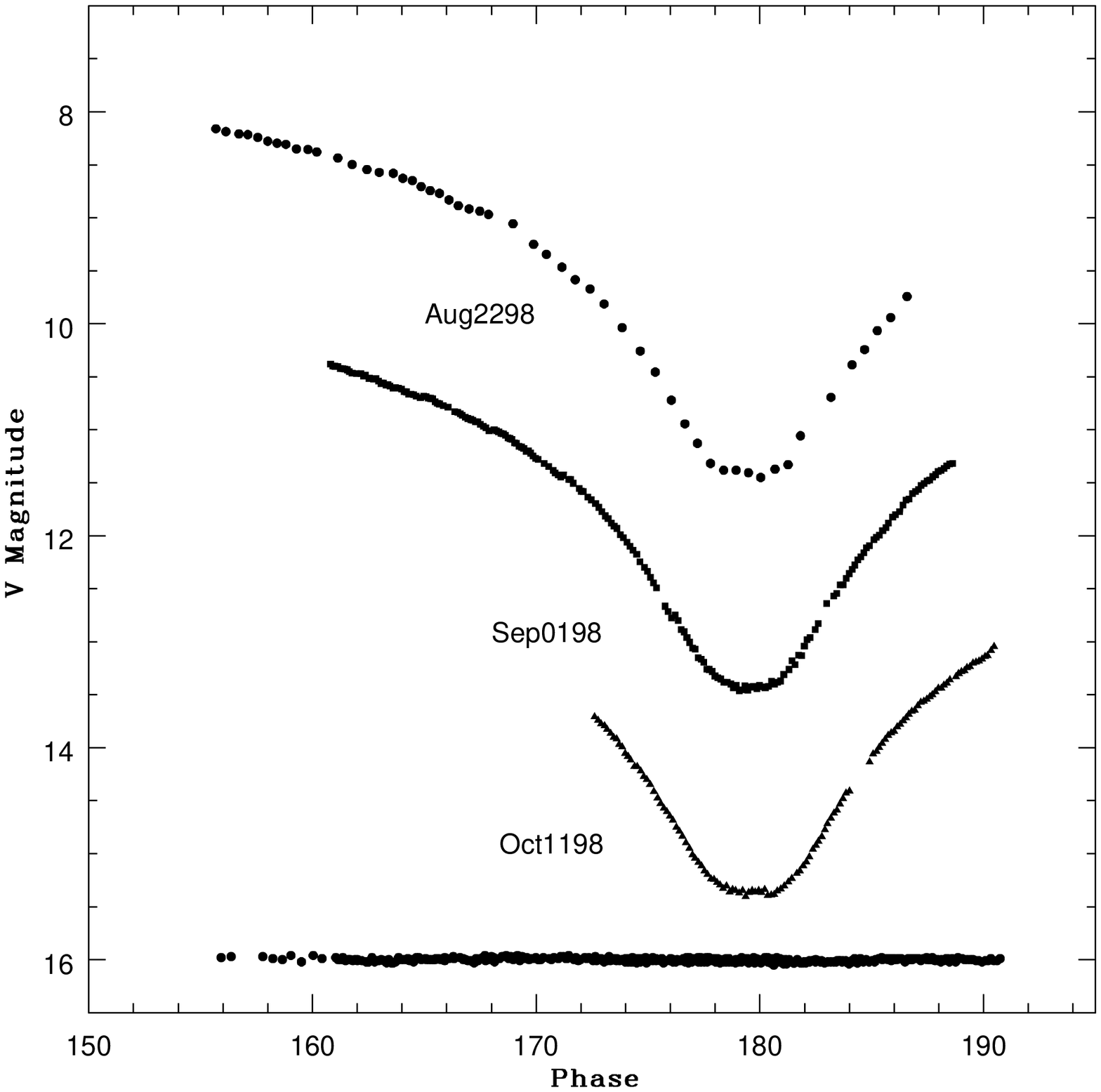}
\figcaption{V light curves for the eclipses observed in 1998 on the indicated
UT dates. August and October data are offset by $-$2 and $+$2 magnitudes 
respectively. Comparison star differences are shown at the bottom.}
\end{figure}

\begin{figure}
\vspace{19.2cm}
\includegraphics{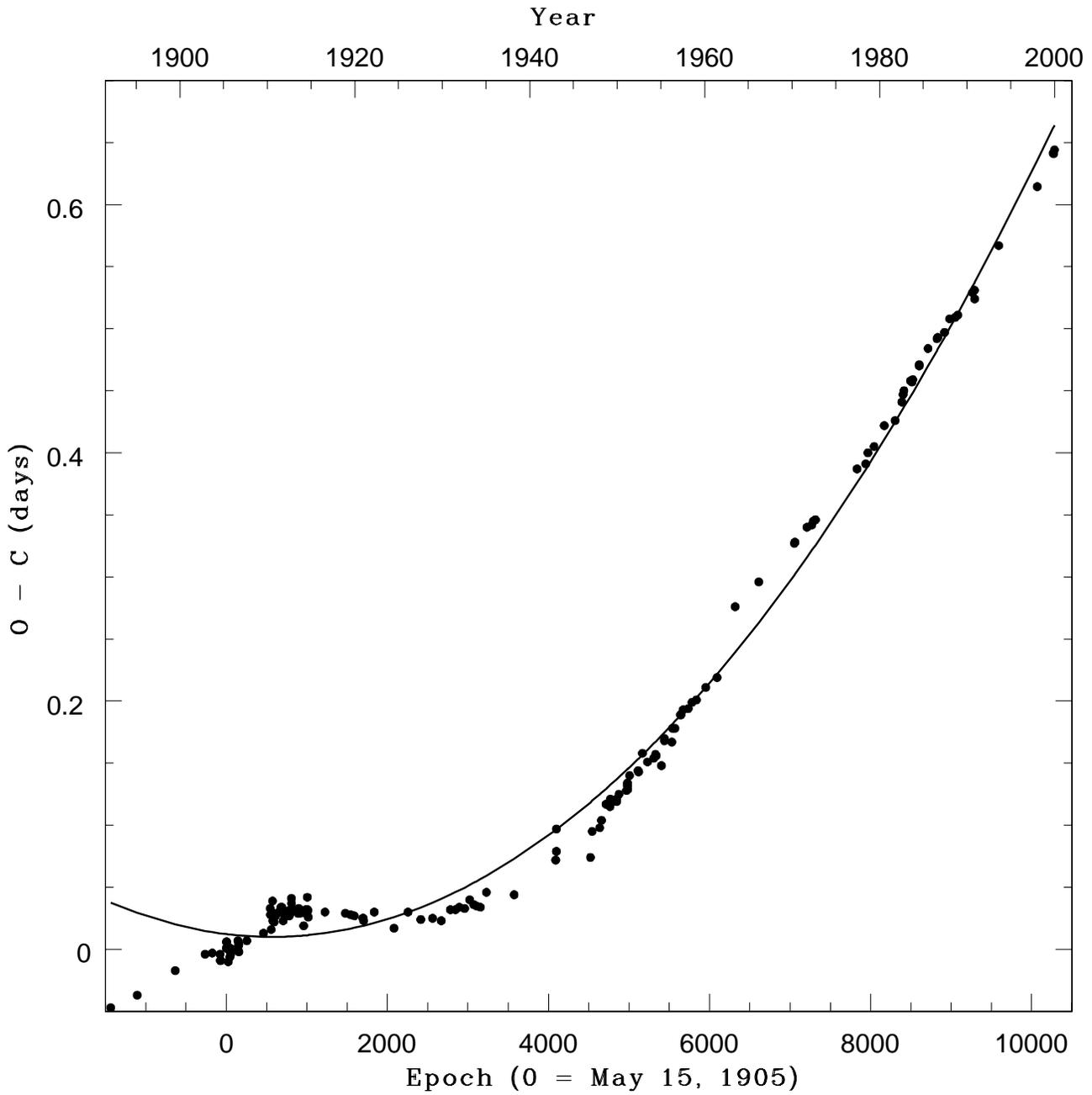}
\figcaption{O$-$C plot for WW Cyg with respect to Graff's ephemeris (equation 2).
The top label shows approximate calendar years of the observations. The solid
line represents the best fit parabola to the data (equation 3). Data for epochs $<$ 
7500 are from \citet{haw72}.}
\end{figure}

\begin{figure}
\vspace{19.2cm}
\includegraphics{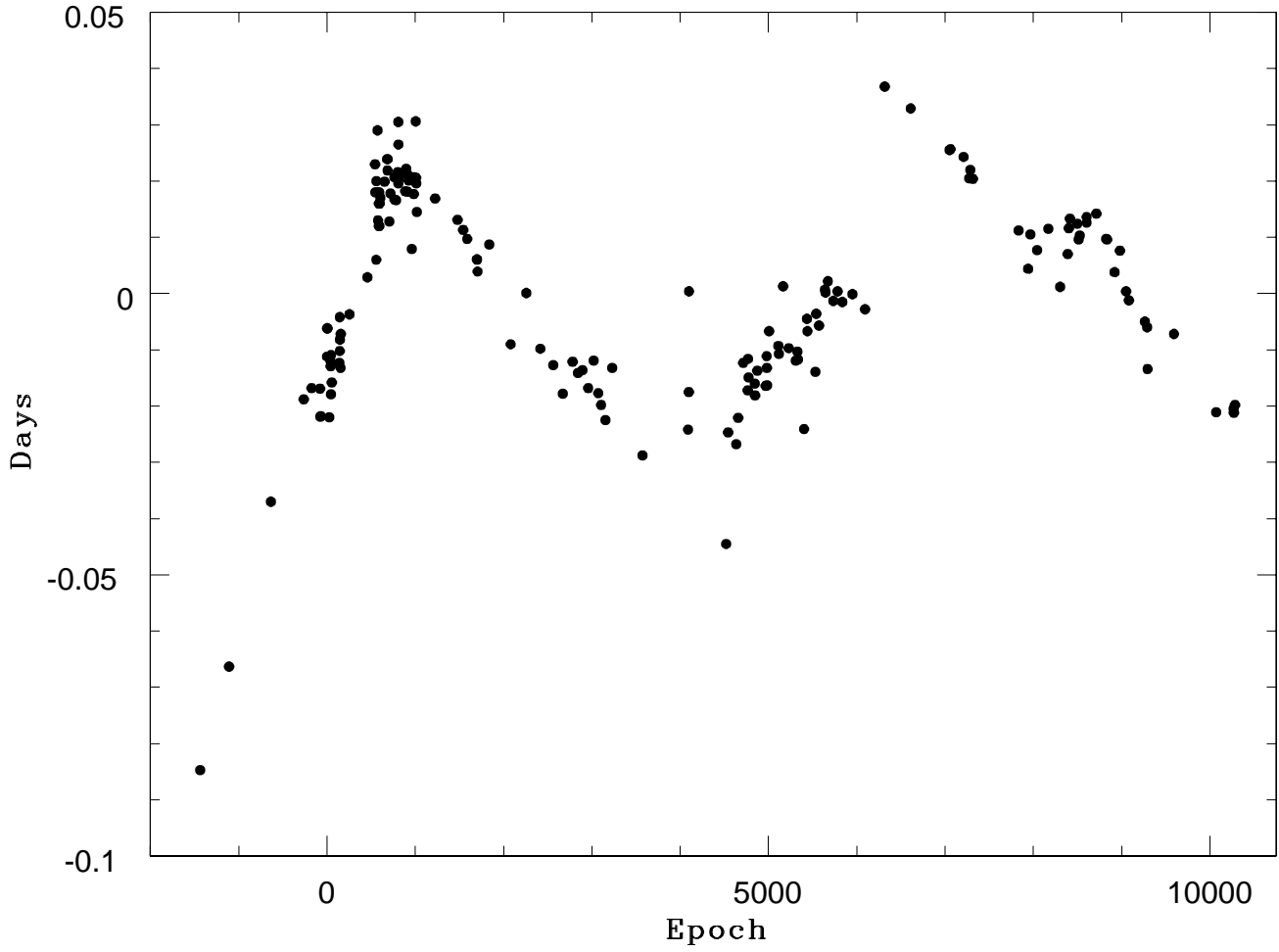}
\figcaption{The O$-$C residuals in days for WW Cyg after subtraction of the best-fit 
parabola given by equation 3.
}
\end{figure}

\begin{figure}
\vspace{19.2cm}
\includegraphics{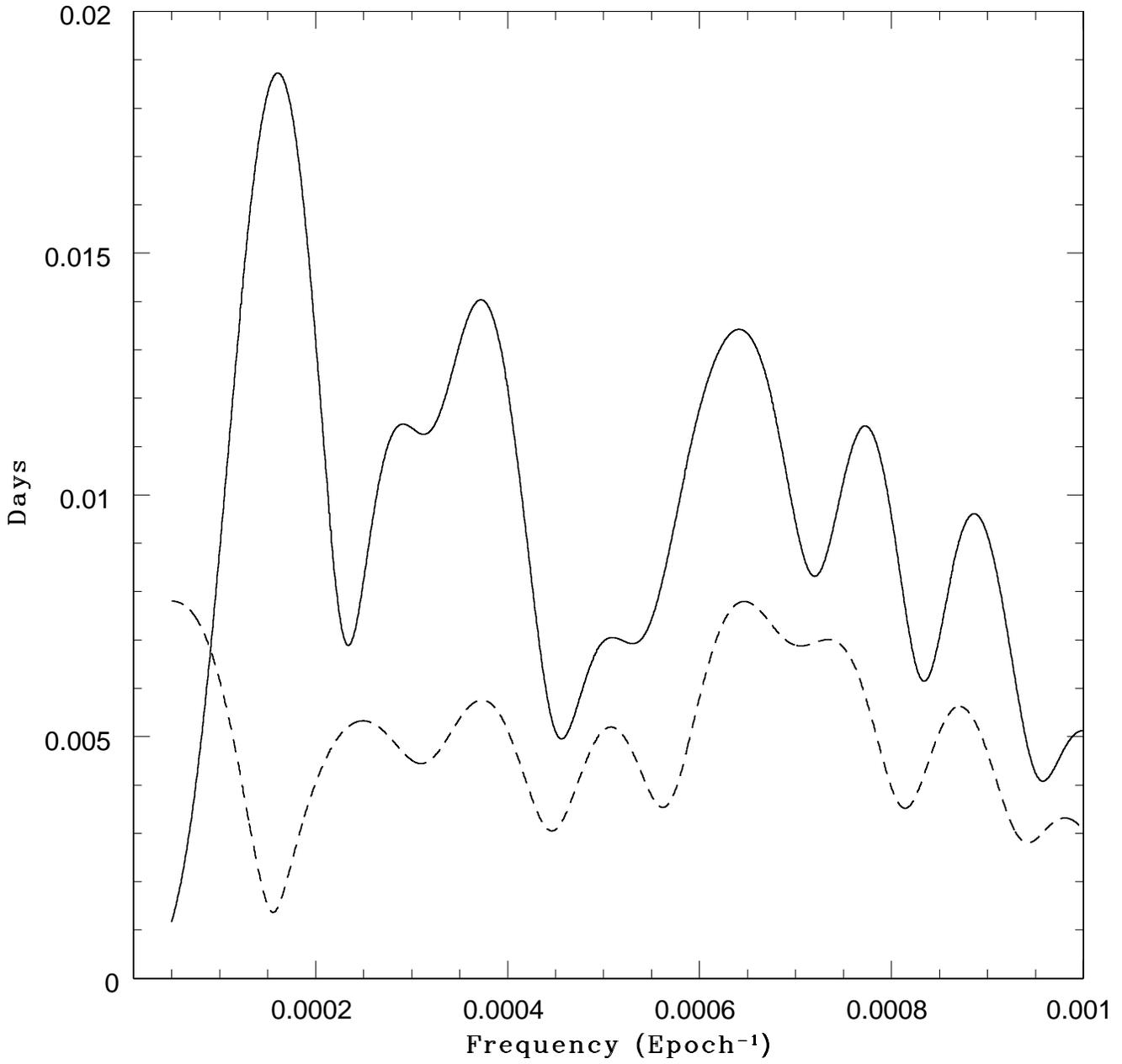}
\figcaption{Fourier transform of the data shown in Figure 3 (solid line). 
 The dashed line is the Fourier transform after subtraction of the waveform in equation 6.}
\end{figure}

\begin{figure}
\vspace{19.2cm}
\includegraphics{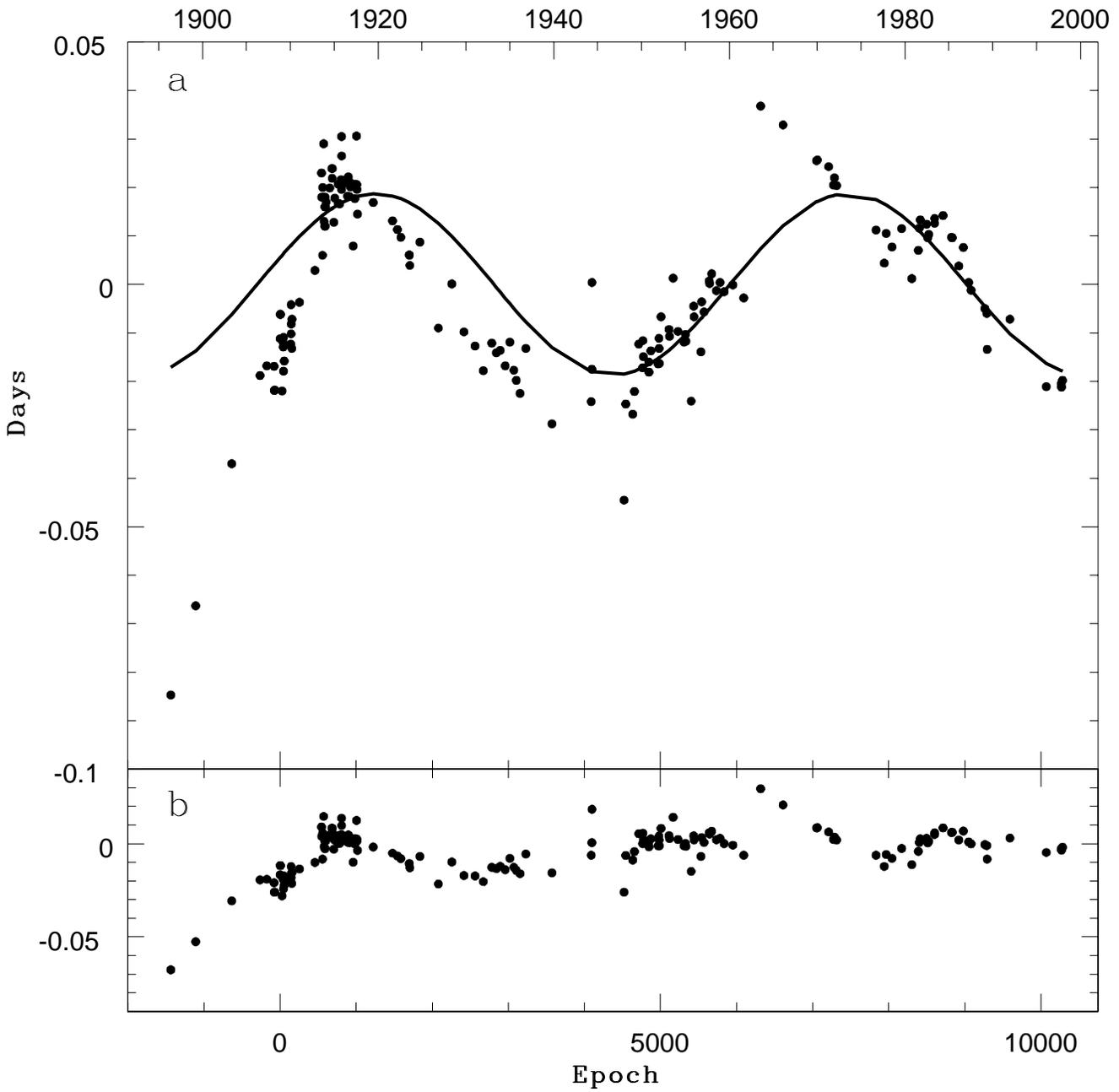}
\figcaption{Figure 5a shows the residuals of fig. 3 with 
equation 6 plotted as the solid line. Figure 5b shows the residuals after
the curve shown in Figure 5a is subtracted from the data. 
Approximate calendar years appear along the top.}
\end{figure}

\begin{figure}
\vspace{19.2cm}
\includegraphics{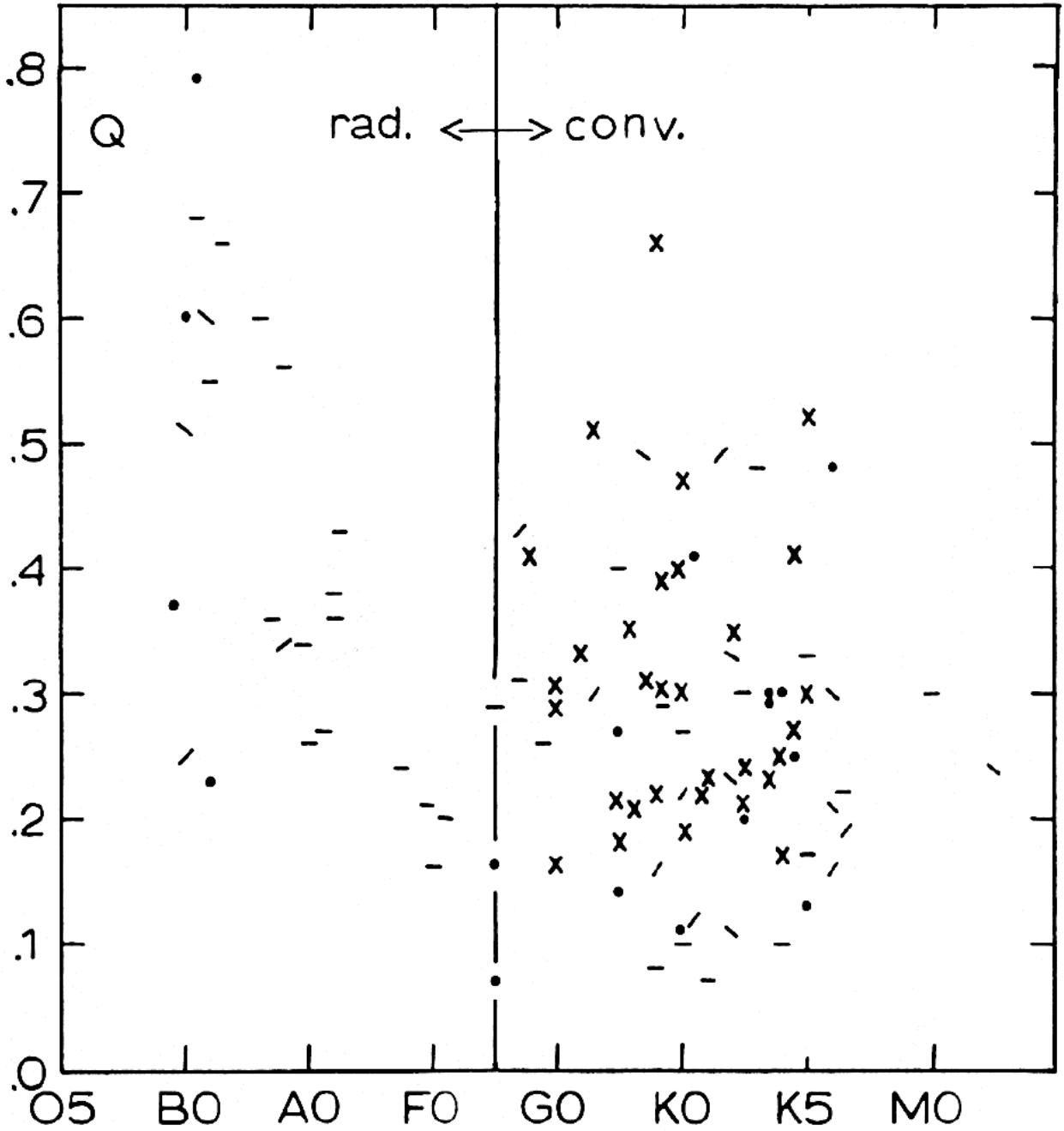}
\figcaption{Plot of mass ratio Q versus secondary spectral type taken from 
\citet{hall89}. A horizontal line indicates no period change noted, 
$/$ indicates a period increase only, $\backslash$  indicates a period 
decrease only, and an X indicates both increases and decreases of the period. 
A $\bullet$ is used for systems for which no conclusion about the period trend 
could be drawn.}
\end{figure}

\begin{figure}
\vspace{19.2cm}
\includegraphics{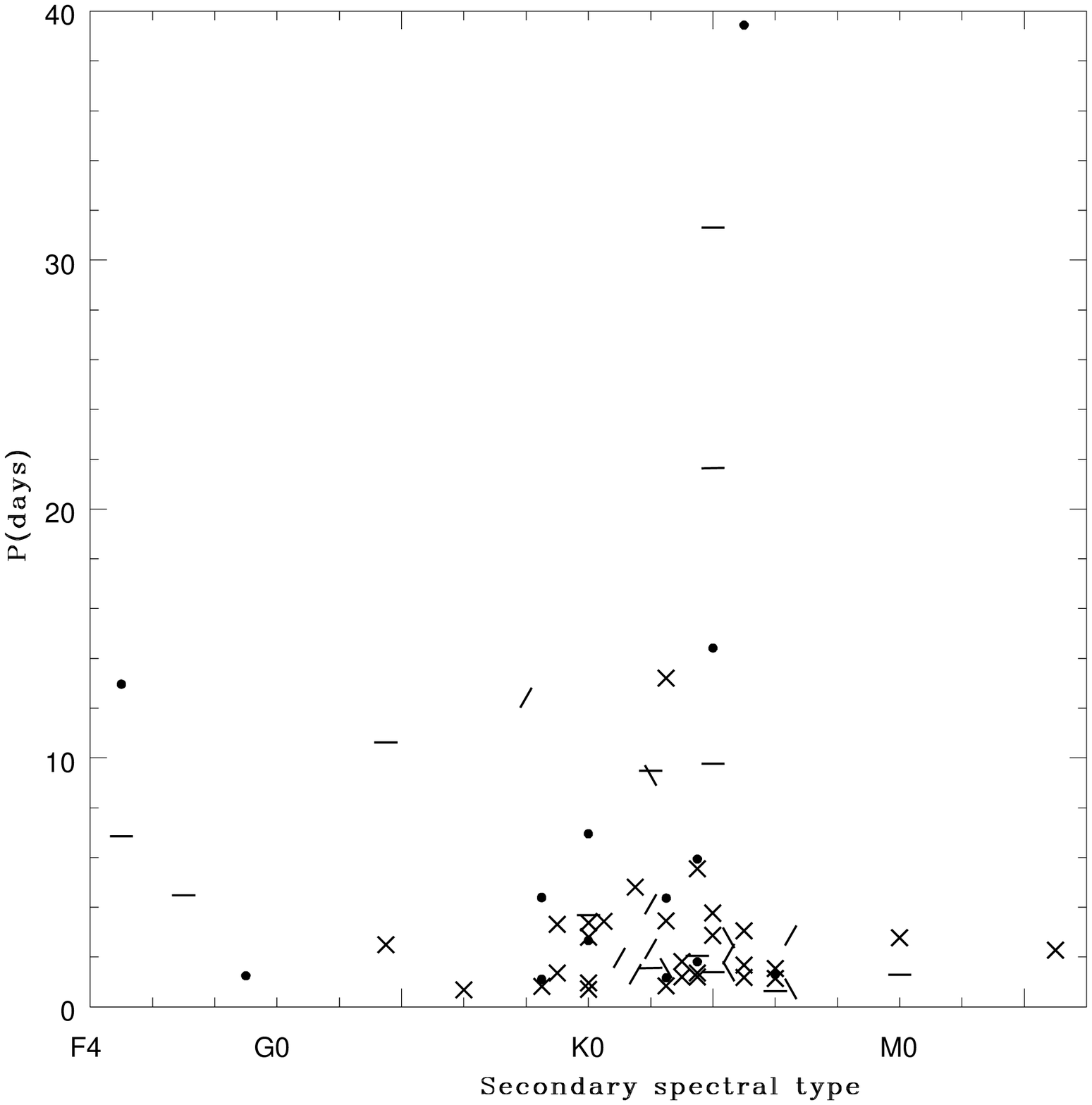}
\figcaption{Plot of binary period P in days versus secondary spectral type for 
the Algols listed in Table 3. The type of period change follows the convention
adopted in Figure 6.}
\end{figure}

\clearpage


\begin{deluxetable}{rcc}
\tablecaption{O{\sc bserved} P{\sc rimary} M{\sc inima of} \ww. \label{tbl-1}}
\tablewidth{0pt}
\tablehead{
\colhead{Epoch of Minimum} & \colhead{Band}   & \colhead{Number of Points} \\
(HJD $-$2450000)
}
\startdata
387.6074 $\pm$ 0.0005\tablenotemark{a} & R & --- \\
1047.8520 $\pm$ 0.0006 & V & 53 \\
1057.8047 $\pm$ 0.0002 & V & 206 \\
1097.6198 $\pm$ 0.0002 & V & 120 \\
\enddata
\tablenotetext{a}{\citet{buck98}}

\end{deluxetable}

\clearpage


\begin{deluxetable}{lrrc}
\tablecaption{O$-$C {\sc data for} \ww\tablenotemark{a}. \label{tbl-2}}
\tablewidth{0pt}
\tablehead{
\colhead{HJD} & \colhead{Epoch}   & \colhead{O$-$C} & \colhead{Ref.}\\
($-$2400000)
}
\startdata
42965.739 & 7315 & 0.346 & 1 \\
43320.734 & 7939 & 0.391 & 1 \\
43413.638 & 7967 & 0.400 & 1 \\
43665.786 & 8043 & 0.405 & 1 \\
44083.831 & 8169 & 0.422 & 1 \\
44531.721 & 8304 & 0.426 & 1 \\
44813.738 & 8389 & 0.441 & 1 \\
44856.874 & 8402 & 0.447 & 1 \\
44896.689 & 8414 & 0.450 & 1 \\
45168.747 & 8496 & 0.458 & 1 \\
45221.829 & 8512 & 0.457 & 1 \\
45261.643 & 8524 & 0.459 & 1 \\
45523.750 & 8603 & 0.470 & 1 \\
45523.751 & 8603 & 0.471 & 1 \\
45888.708 & 8713 & 0.484 & 1 \\
46253.661 & 8823 & 0.492 & 1 \\
46286.839 & 8833 & 0.493 & 1 \\
46568.845 & 8918 & 0.497 & 1 \\
46774.552 & 8980 & 0.508 & 1 \\
47006.790 & 9050 & 0.509 & 1 \\
47109.640 & 9081 & 0.511 & 1 \\
47716.793 & 9264 & 0.529 & 1 \\
47799.737 & 9289 & 0.531 & 1 \\
47809.683 & 9292 & 0.524 & 1 \\
48801.711 & 9591 & 0.567 & 1 \\
50387.6074 & 10069 & 0.6144 & 2 \\
51047.8520 & 10268 & 0.6414 & 3 \\
51057.8047 & 10271 & 0.6411 & 3 \\
51097.6198 & 10283 & 0.6441 & 3 \\
\enddata
\tablenotetext{a}{Based on ephemeris of \citet{gra22}}
\tablerefs{ (1) \citet{bsam95}; (2) \citet{buck98}; (3) This work}
\end{deluxetable}
\clearpage


\begin{deluxetable}{ccccc|ccccc}
\tabletypesize{\scriptsize}
\tablecaption{P{\sc eriod changes of} A{\sc lgols with convective secondaries}. 
\label{tbl-3}}
\tablewidth{0pt}
\tablehead{
\colhead{Star} & \colhead{Sec.} & \colhead{P} & \colhead{Type of} 
& \colhead{Ref.} & \colhead{Star} & \colhead{Sec.} & \colhead{P} & \colhead{Type of} 
& \colhead{Ref.} \\
 & {Sp.T.} & (days) & {$\Delta$}P & & & {Sp.T.} & (days) & {$\Delta$}P }
\startdata
TW And & K2\tablenotemark{a} & 4.1228 & / & 1 & AD Her & K4 & 9.7666 & $-$ & 29 \\ 
XZ And & G9 & 1.3573 & X & 2 & SZ Her & G8-G9 & 0.8181 & X & 1,27 \\ 
RY Aqr & K1 & 1.9666 & / & 3 & UX Her & K6 & 1.5489 & X & 1,28  \\  
KO Aql & K6-K7 & 2.8640 & / & 4 & V338 Her & K6 & 1.3057 & \bul & 1 \\
V346 Aql & G8-G9 & 1.1064 & \bul & 5 & RX Hya & M5 & 2.2816 & X & 30 \\
RW Ara & K2-K3 & 4.3674 & \bul & 6 & TT Hya & K0 & 6.9534 & \bul & 31 \\
IM Aur & F9 & 1.2473 & \bul & 7 & Y Leo & K5 & 1.6861 & X & 1 \\
SU Boo & K2 & 1.5612 & $-$ & 1 & RS Lep & M0 & 1.2885 & $-$ & 32 \\
S Cnc & K2 & 9.4846 & $-$ & 8 & $\delta$ Lib & K2 & 2.3274 & / & 28,33 \\
RZ Cnc & K4III & 21.6430 & $-$ & 9 & RV Oph & K0 & 3.6871 & $-$ & 1 \\
RZ Cas & K3-K4 & 1.1953 & X & 11 & UU Oph & G8-G9 & 4.3968 & \bul & 34 \\
TV Cas & K3 & 1.8126 & X & 12 & DN Ori & F5III\tablenotemark{c} & 12.9663 & \bul & 35 \\
TW Cas & K4-K5 & 1.4283 & \bs & 13 & AQ Peg & K3-K4 & 5.5485 & X & 1 \\
CV Car & K4 & 14.4149 & \bul & 10 & AT Peg & K6 & 1.1461 & X & 36 \\
U Cep & G3-G4 & 2.4930 & X & 1 & AW Peg & G3-G4 & 10.6225 & $-$ & 50 \\
RS Cep & G8III-IV\tablenotemark{b} & 12.4199 & / & 15 & DI Peg & K0 & 0.7118 & X & 37 \\
XY Cep & K4-K5 & 2.7745 & \bs & 16 & $\beta$ Per & K4 & 2.8673 & X & 43 \\
U CrB & K2-K3 & 3.4522 & X & 1 & RT Per & K2-K3 & 0.8494 & X & 1 \\
RW CrB & K6-K7 & 0.7264 & \bs & 17 & RW Per & K2-K3 & 13.1939 & X & 38 \\
UZ Cyg & K4 & 31.3058 & $-$ & 18 & RY Per & F5g & 6.8636 & $-$ & 1 \\
WW Cyg & G9 & 3.3178 & X & 19,20 & Y Psc & K4 & 3.7658 & X & 1 \\
ZZ Cyg & K6 & 0.6286 & $-$ & 1 & U Sge & K0 & 3.3806 & X & 39 \\
KU Cyg & K5III & 39.4393 & \bul & 8 & V505 Sgr & K5 & 1.1829 & X & 40 \\
V548 Cyg & K3-K4 & 1.8053 & \bul & 21 & HU Tau & K3-K4 & 2.0563 & $-$ & 41 \\
W Del & K1-K2 & 4.8060 & X & 1 & RW Tau & M0 & 2.7688 & X & 42 \\
Z Dra & K3-K4 & 1.3574 & X & 1 & X Tri & K0 & 0.9715 & X & 44 \\
TW Dra & K0 & 2.8069 & X & 1 & TX UMa & K5 & 3.0632 & X & 1 \\
AI Dra & K3 & 1.1988 & X & 22 & VV UMa & G6 & 0.6874 & X & 45 \\
S Equ & K0-K1 & 3.4361 & X & 1 & S Vel & K3-K4 & 5.9336 & \bul & 46 \\
AS Eri & K0 & 2.6642 & \bul & 24 &  DL Vir & K1-K2 & 1.3155 & / & 47 \\
AL Gem & K4 & 1.3913 & $-$ & 25 &  RS Vul & F7 & 4.4777 & $-$ & 8 \\ 
RY Gem & K2 & 9.3009 & \bs & 23 & BE Vul & K2-K3 & 1.5520 & \bs & 48 \\
X Gru & K4-K5 & 2.1236 & / & 26 &  V78 $\omega$Cen & K2-K3 & 1.1681 & \bul & 49 \\
\enddata
\tablenotetext{a}{Ref. 19}
\tablenotetext{b}{Ref. 14}
\tablenotetext{c}{Ref. 35}

\tablerefs{ (1) Kreiner 1971; (2) Demircan et al. 1995; (3) Helt 1987;
(4) Hayasaka 1979; (5) Quester 1968; (6) Walter 1976; (7) G$\ddot{\rm u}$lmen et al. 
1985; (8) Kreiner \& Ziolkowski 1978; (9) Olson 1989; (10) Giuricin \& Mardirossian 1981;
(11) Narusawa, Nakamura, \& Yamasaki 1994; (12) Grauer 1977; 
(13) McCook 1971; (14) Plavec \& Dobias 1987; (15) Hall, Cannon \& Rhombs 1973; 
(16) Kreiner \& Tremko 1988; (17) Qian 2000a; (18) Hall \& Woolley 1973; 
(19) Yoon et al. 1994; (20) This work (21) Ertan 1981; (22) Degirmenci et al. 2000; 
(23) Hall \& Stuhlinger 1976; (24) Koch 1960; (25) Koch 1963; (26) Schultz \& Walter 1977;
(27) Baldwin \& Samolyk 1997; (28) Baldwin \& Samolyk 1996; (29) Batten \& Fletcher 1978; 
(30) Qian 2000b; (31) Kulkarni \& Abyankar 1980;
(32) Shenghong et al. 1994; (33) Koch 1962; (34) Taylor 1981; (35) Etzel \& Olson 1995;
(36) Rovithis-Livaniou \& Rovithis 2001; (37) Lu 1992; (38) Mayer 1984; (39) Simon 1997; 
(40) Qian, Liu, \& Yang 1998; (41) Parthasarathy \& Sarma 1980; (42)Frieboes-Conde \& Herczeg 1973; 
(43) Soderhjelm 1980;
(44) Rovithis-Livaniou et al. 2000; (45) Simon 1996; (46) Sister{\' o} 1971; 
(47) Sch$\ddot{\rm o}$effel 1977;
(48) Qian 2001; (49) Sister{\' o}, Fourcade \& Laborde 1969;  
(50) Derman \& Demircan 1992.
}
\end{deluxetable}

\clearpage


\begin{thebibliography}{}
\bibitem[Applegate (1992)]{App92} Applegate, J.H. 1992, ApJ, 385, 621
\bibitem[Applegate \& Patterson (1987)]{App87} Applegate, J.H.,\& Patterson, J. 1987, ApJ, 322, L99
\bibitem[Baldwin \& Samolyk (1995)]{bsam95} Baldwin, M.~E. \& Samolyk, G. 
1995, Observed Minima Timings of Eclipsing Binaries, Number 2 
(Cambridge:AAVSO)
\bibitem[Baldwin \& Samolyk (1996)]{bsam96} Baldwin, M.~E. \& Samolyk, G. 
1996, Observed Minima Timings of Eclipsing Binaries, Number 3 
(Cambridge:AAVSO)
\bibitem[Baldwin \& Samolyk (1997)]{bsam97} Baldwin, M.~E. \& Samolyk, G. 
1997, Observed Minima Timings of Eclipsing Binaries, Number 4 
(Cambridge:AAVSO)
\bibitem[Baliunas et al. (1995)]{bal95}Baliunas, et al. 1995, \apj, 438, 269
\bibitem[Batten \& Fletcher (1978)]{bf78}Batten, A.~H. \& Fletcher, J.~M.
1978, \pasp, 90, 312
\bibitem[Bessell(1990)]{bess90} Bessell, M.~S.\ 1990, \pasp, 
102, 1181 
\bibitem[Borkovits \& Heged$\rm\ddot{u}$s (1996)]{bheg96} Borkovits, T. \& 
Heged$\rm\ddot{u}$s, T. 1996, A\&AS, 120, 63
\bibitem[Buckner et al. (1998)]{buck98} Buckner, M., Nellermoe, B., \& Mutel, 
R. 1998, IBVS, 4559
\bibitem[Chambliss (1992)]{cham92}Chambliss, C.R. 1992 PASP, 104, 663
\bibitem[DeCampli and Baliunas (1979)]{deb79} DeCampli, W. M. \& Baliunas, 
S. L. 1979, ApJ, 230, 815
\bibitem[Degirmenci et al. (2000)]{deg00}Degirmenci, $\ddot{\rm O}$.~L.,
G$\ddot{\rm u}$lmen, $\ddot{\rm O}$, 
Sezer, C., Erdem, A., \& Devlen, A. 2000, \aap, 363, 244
\bibitem[Demircan et al. (1995)]{dem95} Demircan, O., Akalin, A.,
Selam, S., \& Mueyesseroglu, Z. 1995, A\&AS, 114, 167 
\bibitem[Derman \& Demircan (1992)]{derdem92}Derman, E. \& Demircan, 
O. 1992, \apss, 189, 309
\bibitem[Ertan (1981)]{ert81}Ertan, A.~Y. 1981, ApSS, 77, 391
\bibitem[Etzel \& Olson (1995)]{eo95}Etzel, P.~B. \& Olson, E.~C. 1995, 
\aj, 110, 1809
\bibitem[Frieboes-Conde \& Herczeg (1973)]{fch73} Frieboes-Conde, H. \& 
Herczeg, T. 1973, A\&AS, 12, 1
\bibitem[Giuricin \& Mardirossian (1981)]{gmar81}Giuricin, G. \&  Mardirossian,
F. 1981, A\&AS, 45, 85
\bibitem[Giuricin, Mardirossian, \& Mezzetti (1983)]{giu83} Giuricin, G., 
Mardirossian, F., \& Mezzetti, M. 1983, \apjs, 52,35
\bibitem[Graff (1922)]{gra22}Graff, K. 1922, AN, 217, 349
\bibitem[Grauer, et al. (1977)]{grau77}Grauer, A.~D., McCall, J., Reaves, 
L.~C., \& Tribble, T.~L. 1977, \aj, 82, 740
\bibitem[G$\ddot{\rm u}$lmen,  et al. (1985)]{gul85}G$\ddot{\rm 
u}$lmen, $\ddot{\rm O}$, Sezer, C. \& G$\ddot{\rm u}$d$\ddot{\rm u}$r, N. 
1985, A\&AS, 60, 389
\bibitem[Hall (1976)]{hall76}Hall,D.~S. 1976, in IAU Colloq. No. 29, 
Multiple Periodic Variable Stars, ed. W.~S. Fitch (Dordrecht:Reidel), 287
\bibitem[Hall (1989)]{hall89}Hall, D.~S. 1989, Space Sci. Rev., 50, 219
\bibitem[Hall (1991)]{hall91}Hall, D.~S. 1991, ApJ, 380, L85
\bibitem[Hall, Cannon, \& Rhombs (1973)]{hcr73}Hall, D.~S., Cannon III, R.~O., \& 
Rhombs, C.~G. 1973, \pasp, 85, 420
\bibitem[Hall \& Kreiner (1980)]{hk80} Hall, D.~S. \& Kreiner, J.M. 1980, Acta
Astron., 30, 287
\bibitem[Hall \& Stuhlinger (1976)]{hs76} Hall, D.~S. \& Stuhlinger, T., 
 1976, Acta Astron., 26, 109
\bibitem[Hall \& Wawrukiewicz (1972)]{haw72} Hall, D.~S. and Wawrukiewicz, 
A.~S. 1972, \pasp, 84, 541
\bibitem[Hall \& Woolley (1973)]{hwoo73} Hall, D.~S. \& Woolley, K.~S.,
1973, \pasp, 85, 618
\bibitem[Hayasaka (1979)]{hay79}Hayasaka, T. 1979 PASJ, 31, 271
\bibitem[Helt (1987)]{helt87}Helt, B.~E. 1987, \aap, 172, 155
\bibitem[Hobart et al. (1994)]{hob94} Hobart, M.~A., Pena, J.~H., Peniche, R.,
 Rodriguez, E., Garrido, R., Rios-Berumen, M., Rios-Herrera, M., 
\& Lopez-Cruz, O. 1994 Revista Mexicana de Astronomia y Astrofisica, 28, 111
\bibitem[Koch (1960)]{koch60}Koch, R.~H. 1960, \aj, 65, 139
\bibitem[Koch (1962)]{koch62}Koch, R.~H. 1962, \aj, 67, 130
\bibitem[Koch (1963)]{koch63}Koch, R.~H. 1963, \aj, 68, 785
\bibitem[Kopal (1978)]{kop78} Kopal, Z. 1978, Dynamics of close binary systems
(Dordrecht:D. Reidel)
\bibitem[Kreiner (1971)]{krei71} Kreiner, J.~M. 1971, Acta. Astron., 21, 365
\bibitem[Kreiner \& Tremko (1988)]{kt88}Kreiner, J.~M. \& Tremko, J. 1988,
BAICz, 39, 73
\bibitem[Kreiner \& Ziolkowski (1978)]{kz78}Kreiner, J.~M. \& Ziolkowski,
J. 1978, Acta Astron., 28, 497
\bibitem[Kulkarni \& Abyankar (1980)]{kab80}Kulkarni, A.~G. \& Abyankar, 
K.~D. 1980, \apss, 67, 205 
\bibitem[Kwee \& van Woerden (1956)]{kvw56} Kwee, K.~K., and van Woerden, H.  
1956, \bain, 12, 357
\bibitem[Kwee \& van Woerden (1958)]{kvw58} Kwee, K.~K., and van Woerden, H.  
1958, \bain, 485, 131
\bibitem[Lanza \& Rodon$\rm\grave{o}$ (1999)]{lan99} Lanza, A.~F. \& 
Rodon$\rm\grave{o}$, M. 1999, \aap, 349, 887
\bibitem[Lanza, Rodon$\rm\grave{o}$, and Rosner (1998)]{Lan98} Lanza A.F., Rodono M., 
Rosner, R. 1998, MNRAS, 296,893
\bibitem[Lu(1992)]{lu92} Lu, W.\ 1992, Acta Astronomica, 42, 
73 
\bibitem[Mallama (1982)]{mal82}Mallama, A.~D. 1982, Communications of the IAPPP, 
7, 14
\bibitem[Marsh \& Pringle (1990)]{Mar90}Marsh, T.R. \& Pringle, J.E. 1990, 
\apj, 365, 677
\bibitem[Massey \& Jacoby (1992)]{maj92} Massey, P. \& Jacoby, G.~H. 1992 in 
ASP Conf. Ser. 23, Astronomical CCD Observing and Reduction Techniques, ed. 
S.~B. Howell (San Francisco:ASP), 240
\bibitem[Mayer(1984)]{may84} Mayer, P.\ 1984, BAICz, 35, 180 
\bibitem[McCook (1971)]{mcc71}McCook, G.~P. 1971, \aj, 76, 449
\bibitem[Narusawa, Nakamura, \& Yamasaki (1994)]{nny94}Narusawa, S., Nakamura, 
Y. \& Yamasaki, A. 1994, \aj, 107. 1141 
\bibitem[Olson (1989)]{ols89}Olson, E.~C. 1989 \aj, 98, 1002
\bibitem[Parthasarathy \& Sarma(1980)]{psar80} Parthasarathy, 
M.\ \& Sarma, M.\ B.\ K.\ 1980, \apss, 72, 477 
\bibitem[Parker (1979)]{par79} Parker, E.N. 1979, Cosmical magnetic fields 
(Oxford:Clarendon Press)
\bibitem[Plavec \& Dobias (1987)]{pd87}Plavec, M.~J., \& Dobias, J.~J. 1987, 
\aj, 93, 171
\bibitem[Press et al. (1992)]{pre92} Press, W.~H., Teukolsky, S.~A., 
Vetterling, W.~T., \& Flannery, B.~P. 1992, Numerical Recipes in Fortran 77 
(2nd ed.; Cambridge: Cambridge University Press)
\bibitem[Quester (1968)]{ques68}Quester, W. 1968, IBVS, 283
\bibitem[Qian (2000a)]{qian00a}Qian, S. 2000a, \aj, 119, 901
\bibitem[Qian (2000b)]{qian00b}Qian, S. 2000b, A\&AS, 146, 377
\bibitem[Qian(2001)]{qian01} Qian, S.\ 2001, \aj, 121, 1614 
\bibitem[Qian, Liu, \& Yang(1998)]{qly98} Qian, S., Liu, Q., 
\& Yang, Y.\ 1998, Publications of the Yunnan Observatory, 75, 1 
\bibitem[Richards(1990)]{rich90} Richards, M.\ T.\ 1990, \apj, 
350, 372 
\bibitem[Richards \& Albright(1993)]{rab93} Richards, M.\ T.\ 
\& Albright, G.\ E.\ 1993, \apjs, 88 199 
\bibitem[Rovithis-Livaniou et al. (2000)]{rkr00} Rovithis-Livaniou, H., 
Kranidiotis, A.\ N., Rovithis, P., \& Athanassiades, G.\ 2000, \aap, 354, 
904 
\bibitem[Rovithis-Livaniou \& Rovithis (2001)]{rlr01}Rovithis-Livaniou, H. 
\& Rovithis, P., 2001 in ASP Conf. Ser. 229, Evolution of Binary and Multiple Star
Systems, ed. Ph. Podsiadlowski, S, Rappaport, A.~R. King, F. D'Antona \& 
L. Burderi (San Francisco:ASP) 2001
\bibitem[Sch$\ddot{\rm o}$effel(1977)]{schof77} 
Sch$\ddot{\rm o}$effel, E.\ 1977, \aap, 61, 107 
\bibitem[Schultz \& Walter (1977)]{sw77} Schultz, E. \& Walter, K.  1977, 
A\&AS, 29, 51
\bibitem[Shenghong et al.(1994)]{sqy94} Shenghong, G., 
Quingyao, L., Yulan, Y., Bi, W., \& Zhankui, H.\ 1994, IBVS, 4060, 1
\bibitem[Simon(1996)]{sim96} Simon, V.\ 1996, \aap, 311, 915 
\bibitem[Simon(1997)]{sim97} Simon, V.\ 1997, \aap, 327, 1087 
\bibitem[Sister{\' o}(1971)]{sis71} Sister{\' o}, R.\ F.\ 
1971, BAICz, 22, 188 
\bibitem[Sister{\' o}, Fourcade \& Laborde (1969)]{sfl69} Sister{\' o}, R.\ F.\, 
Fourcade, C.~R. \& Laborde, J.~R. 1969, IBVS, 402 
\bibitem[Soderhjelm(1980)]{sod80} Soderhjelm, S.\ 1980, \aap, 
89, 100 
\bibitem[Struve (1946)]{Str46} Struve, O. 1946, \apj, 104, 253
\bibitem[Taylor (1981)]{tay81}Taylor, M. 1981, JAAVSO, 10, 83
\bibitem[Tout \& Hall(1991)]{th91} Tout, C.~A.~\& Hall, 
D.~S.\ 1991, \mnras, 253, 9 
\bibitem[Yoon et al. (1994)]{yoon94} Yoon, T.~S., Honeycutt, R.~K.,
Kaitchuck, R.~H., and Schlegel, E.~M. 1994, \pasp, 106, 239
\bibitem[Walter (1976)]{wal76}Walter, K. 1976, A\&AS, 26, 227
\bibitem[Warner (1988)]{war88} Warner, B.~W. 1988, Nature, 336, 129

\end{thebibliography}
\end{document}